\documentclass[aps,nofootinbib,preprintnumbers,twocolumn,showpacs,eqsecnum]{revtex4}

\usepackage{citesort,latexsym,ifthen,graphics,color}

\newcommand{\ie}{\emph{i.e.}}

\newcommand{\aka}{\emph{a.k.a.}}

\newcommand{\wrt}{with respect to}

\newcommand{\lhs}{left-hand side}
\newcommand{\rhs}{right-hand side}

\newcommand{\be}{\begin{equation}}
\newcommand{\ee}{\end{equation}}
\newcommand{\bea}{\begin{eqnarray}}
\newcommand{\eea}{\end{eqnarray}}
\newcommand{\beas}{\begin{eqnarray*}}
\newcommand{\eeas}{\end{eqnarray*}}

\newcommand{\bear}{\begin{array}{l}}
\newcommand{\eear}{\end{array}}

\newcommand{\bcf}{\begin{figure}}
\newcommand{\ecf}{\end{figure}}

\newcommand{\bct}{\begin{table}}
\newcommand{\ect}{\end{table}}

\def\eq#1{(\ref{#1})}

\def\Eqn#1{Equation~(\ref{#1})}

\def\eqs#1#2{(\ref{#1}) and~(\ref{#2})}

\def\sec#1{section~\ref{#1}}
\def\Sec#1{Section~\ref{#1}}

\def\fig#1{figure~\ref{#1}}
\def\Fig#1{Figure~\ref{#1}}
\def\figs#1#2{figures~\ref{#1} and~\ref{#2}}

\def\Q{{\cal Q}}

\newcommand{\nCr}[2]{ \left.\right.^{#1}\!C_{#2}\ }
\newcommand{\Or}{\mathrm{O}}

\newcommand{\OTower}{
	\begin{array}{c}
	\vspace{1ex}
		\TopO
	\\
		\SumVertex
	\end{array}
}

\def\hS{\hat{S}}

\def\one{\hbox{1\kern-.8mm l}}
\def\str{\mathrm{str}}
\def\tr{\mathrm{tr}}

\newcommand{\Op}[1]{\Or (p^{#1})}

\newcommand{\ODInt}[1]{\int \!\! d#1}

\newcommand{\Int}[1]{\int \!\! d^D \! #1 \,}
\newcommand{\IntB}[1]{\int \!\! \frac{d^D \! #1}{(2\pi)^D} \,}

\newcommand{\volume}[1]{d^D \! #1 \,}

\newcommand{\der}[2]{\ensuremath{\frac{d #1}{d #2}}}
\newcommand{\pder}[2]{\ensuremath{\frac{\partial #1}{\partial #2}}}

\newcommand{\order}[1]{\Or ( #1 )}
\newcommand{\hf}{\frac{1}{2}}

\newcommand{\GIO}{\mathcal{O}}
\newcommand{\fields}{\Phi}
\newcommand{\expectation}[1]{\left\langle #1 \right\rangle}
\newcommand{\measure}[1]{\mathcal{D} #1 \, }
\newcommand{\WL}[1]{\phi(#1)}
\newcommand{\EWL}[1]{W(#1)}
\newcommand{\REWL}[1]{W_R(#1)}
\newcommand{\Loop}{\Gamma}
\newcommand{\length}[1]{l(#1)}
\newcommand{\cusps}{\theta}
\newcommand{\ints}{\vartheta}

\newcommand{\flow}{\Lambda \partial_\Lambda}
\newcommand{\totalflow}{\Lambda \der{}{\Lambda}}
\newcommand{\flowConstAl}{\Lambda \partial_\Lambda|_\alpha}

\newcommand{\dec}[3][0]{\ensuremath{\left[ #2 \hspace{#1em} \right]^{#3}}}
\newcommand{\DummyKernel}{\ensuremath{\stackrel{\bullet}{\mbox{\rule{1cm}{.2mm}}}}}
\newcommand{\norm}{\ensuremath{\Upsilon}}

\newcommand{\GR}{\cdeps{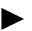}}
\newcommand{\GRk}{\rhd}
\newcommand{\GRkpr}{>}

\newcommand{\bigdot}[1]{\stackrel{\bullet}{#1}}

\newlength{\VertexWidth}

\newcommand{\Vertex}[1]{
\ensuremath{
	\begin{array}{c}
	\settowidth{\VertexWidth}{$#1$}
	\setlength{\unitlength}{1.2\VertexWidth}
	\begin{picture}(1,1)(-0.5,-0.5)
		\put(0,0){\circle{1}}
		\put(-0.4,-0.1){$#1$}
	\end{picture}
	\end{array}
}
}

\newcommand{\SVertex}{
\ensuremath{
	\begin{array}{c}
	\settowidth{\VertexWidth}{1em}
	\setlength{\unitlength}{\VertexWidth}
	\begin{picture}(1,1)(-0.5,-0.5)
		\put(0,0){\circle{1}}
		\put(-0.2,-0.2){$S$}
	\end{picture}
	\end{array}
}
}

\newcommand{\OVertex}[1]{
\ensuremath{
	\begin{array}{c}
	\settowidth{\VertexWidth}{$#1$}
	\setlength{\unitlength}{1.2\VertexWidth}
	\begin{picture}(1,1)(0,0)
		\put(0,0){\framebox(1,1){$#1$}}
	\end{picture}
	\end{array}
}
}

\newcommand{\SumVertex}{
	\Vertex{n_s, j}
}

\newcommand{\TopO}{
	\OVertex{v^{j}}
}

\newcommand{\cdeps}[1]{\ensuremath{\begin{array}{c}\includegraphics{#1.eps} \end{array}}} 
\newcommand{\cd}[1]{\ensuremath{\begin{array}{c}\includegraphics{#1.ps} \end{array}}}

\newcommand{\sco}[3][0]{
	\begin{array}{c}
	\vspace{#1ex}
		#2 
	\\
		#3
	\end{array}
}

\newcommand{\jhep}[3]{{JHEP} #1 (#2) #3}
\newcommand{\NuclPhys}[4]{{Nucl.\ Phys.\ }\textbf{#1 #2} (#3) #4}
\newcommand{\PhysRev}[4]{{Phys.\ Rev.\ }\textbf{#1 #2} (#3) #4}
\newcommand{\IntJModPhys}[4]{{Int.\ J.\ Mod.\ Phys.\ }\textbf{#1 #2} (#3) #4}

\newcommand{\PhysRept}[3]{{Phys.\ Rept.\ }\textbf{#1} (#2) #3}

\newcommand{\AnnPhys}[3]{{Ann.\ Phys.\ }\textbf{#1} (#2) #3}
\newcommand{\PhysLett}[4]{{Phys.\ Lett.\ }\textbf{#1 #2} (#3) #4}
\newcommand{\ProgTheorPhys}[3]{{Prog.\ Theor.\ Phys.\ }\textbf{#1} (#2) #3}
\newcommand{\ProgTheorPhysS}[3]{{Prog.\ Theor.\ Phys.\ Suppl.\ }\textbf{#1} (#2) #3}

\newcommand{\CEurJPhys}[3]{{Central Eur.\ J.\ Phys.\ }\textbf{#1} (#2) #3}

\newcommand{\arxiv}[1]{#1}
\newcommand{\hepth}[1]{hep-th/#1}
\newcommand{\hepph}[1]{hep-ph/#1}

\newcommand{\condmat}[1]{cond-mat/#1}
\newcommand{\CommMathPhys}[3]{{Comm.\ Math.\ Phys.\ }\textbf{#1}, #2 (#3)}
\newcommand{\RevModPhys}[3]{{Rev.\ Mod.\ Phys.\ }\textbf{#1} (#2) #3}
\newcommand{\Polon}[3]{Acta Phys.\ Polon \textbf{#1} (#2) #3}
\newcommand{\jphysa}[3]{J.\ Phys.\ {\bf A}: Math.\ Gen.\ #1 (#2) #3}

\newcommand{\http}[1]{http://#1}

\begin{document}

\title{General Computations Without Fixing the Gauge}

\author{
	Oliver J.~Rosten
}

\affiliation{
Dublin Institute for Advanced Studies, 10 Burlington Road, Dublin 4, Ireland
}
\email{orosten@stp.dias.ie}

\begin{abstract}
	Within the framework of a manifestly gauge invariant
	exact renormalization group for $SU(N)$ Yang-Mills, 
	we derive a simple expression 
	for  the expectation value of an
	arbitrary gauge invariant operator.
	We illustrate the use of this formula
	by computing the $\Or (g^2)$
	correction to the rectangular,
	Euclidean Wilson loop
	with sides $T \gg L$. 
	The standard result is trivially
	obtained, directly in the continuum, for the
	first time without fixing the gauge.
	We comment on possible future applications of the formalism.
\end{abstract}

\preprint{SHEP 06-15, DIAS-STP-06-19}
\maketitle

\tableofcontents

\section{Introduction}
\label{sec:Intro}

The necessity of gauge fixing in order to compute
in Yang-Mills theories is, in common lore,
practically taken for granted and, for 
perturbative calculations,
generally considered obligatory.
This point of view is lent considerable weight
both by Feynman's unitarity argument for
the existence of Faddeev-Popov ghosts~\cite{Feynman}
and by the elegance and power of the resulting 
BRST symmetries~\cite{BRST}.

However, there is no reason in principle
why gauge invariant quantities, as opposed
to Green's functions in gauge fixed formulations, 
cannot be computed in a manifestly gauge invariant manner. 
Indeed, in a non-perturbative context, this is 
routinely exploited
on the lattice, where calculations can be
performed without gauge fixing.
Excitingly, in a series of 
works~\cite{u1,ym,ymi,ymii,sunn,Morris:2000jj,SU(N|N),scalar1,aprop,scalar2,quarks,giqed,Thesis,mgierg1,mgierg2,Primer,RG2005,mgiuc}, 
a formalism has
been developed which allows manifestly
gauge invariant computations to be performed
directly in the continuum.

The benefits of this scheme are numerous.
The gauge field is protected from field strength renormalization and
the Ward identities take a particularly simple form
since the Wilsonian effective action is built only from
gauge invariant combinations of the covariant derivative,
even at the quantum level~\cite{ymi}. 
In the non-perturbative domain, the difficult
technical issue of
Gribov copies~\cite{Gribov} is entirely avoided.
Furthermore, it should be possible to make
statements about phenomena such as confinement
in a completely gauge independent manner, and
it is surely this which gives a manifestly
gauge invariant scheme much of its appeal. 

The framework developed in~\cite{u1,ym,ymi,ymii,sunn,Morris:2000jj,SU(N|N),scalar1,aprop,scalar2,quarks,giqed,Thesis,mgierg1,mgierg2,Primer,RG2005,mgiuc}
is based on the exact renormalization group (ERG), the continuum
version of Wilson's RG~\cite{Wil,W&H,Pol}.
The essential physical idea behind this approach is
that of integrating out degrees of freedom
between the bare scale of the quantum theory and
some effective scale, $\Lambda$. The effects
of these modes are encoded in the Wilsonian effective action, 
$S_\Lambda$, which describes the physics of the theory
in terms of parameters relevant to the effective scale.

The possibility of constructing manifestly gauge invariant
ERGs arises, fundamentally, from the huge freedom inherent in the 
approach~\cite{jose}. For any given quantum field theory,
there are an infinite number of ERGs corresponding
to the infinite number of different ways in which the
high energy degrees of freedom can be averaged over 
(the continuum version of blocking on the 
lattice)~\cite{Morris:2000jj,jose,mgierg1}. In Yang-Mills theory,
an infinite subset of these schemes allow the
computation of the gauge invariant Wilsonian
effective action, without fixing the gauge.\footnote{
In practise we further specialize to those ERGs
which allow convenient renormalization
to any loop order~\cite{mgierg1,mgierg2,Thesis}. 
}

Central to the ERG methodology is the ERG equation,
which determines how the Wilsonian effective action
changes under infinitesimal changes of the scale.
Part of the reason for the considerable amount of
work put into adapting the ERG for Yang-Mills
(see~\cite{Pawlowski:2005xe} for a summary of the various approaches)
is that the ERG equation, 
by relating physics at different
scales, provides 
access to the
low energy dynamics of the theory.
Indeed, more generally, 
the ERG has proven itself to be a flexible
and powerful tool for studying 
both perturbative and non-perturbative problems
in a range of field theories 
(see~\cite{Fisher:1998kv,TRM-elements,Aoki:2000wm,Litim:1998nf,Berges:2000ew,Bagnuls:2000ae,Polonyi:2001se,Salmhofer:2001tr,Delamotte:2003dw} 
for reviews). 
A particular advantage conferred by the ERG is
that renormalization is built in:
solutions to the flow equation (in pretty much any approximation scheme),
from which physics can be extracted,
are naturally phrased directly in terms of renormalized parameters.
It is thus clear that
a manifestly gauge invariant formalism, based
on the ERG, has considerable potential. Furthermore,
an interesting link between this formalism and
the AdS/CFT correspondence has recently been 
made~\cite{EMR}.

The majority of the work into the scheme employed in
this paper
has  focused on constructing and testing
the formalism,
culminating in the successful reproduction of
the $SU(N)$ Yang-Mills two-loop 
$\beta$-function~\cite{Thesis,mgierg2}. Subsequent
to this, the powerful diagrammatic techniques
developed to facilitate this calculation
have been refined and applied in the context
of $\beta$-function coefficients at arbitrary
loop order~\cite{Primer,RG2005,mgiuc}. 
These substantial works have paved the way for more
general application of the formalism;
in this paper, we 
describe how to compute the expectation
values of gauge invariant operators, without
fixing the gauge, and illustrate the formalism
with a very simple computation of the $\Or (g^2)$
correction to the rectangular, Euclidean Wilson loop with 
sides $T \gg L$.
[There have been attempts to compute the perturbative
corrections to this Wilson loop in (gauge fixed) ERG studies,
in the past. In particular, using the axial gauge
flow equation proposed by~\cite{Simionato}, it was found
in~\cite{Panza}
that, whilst the $\Or (g^2)$ result could be correctly
reproduced, the formalism failed at $\Or (g^4)$. However,
the flow equation of~\cite{Simionato}
is not a flow equation in the Wilsonian sense: the implementation
of the cutoff, $\Lambda$, is not sufficient to regularize
the theory, and dimensional regularization has to be employed
as well. This is the reason for the negative result of~\cite{Panza};
as recognized in~\cite{axial}, properly Wilsonian axial
gauge flow equations can be constructed, which work perfectly
well. In the formalism used in this paper, the above issues
never arise, since the implementation of the (gauge invariant) cutoff
is sufficient to regularize the theory, as proven in~\cite{SU(N|N)}.]

The outline of this paper is as follows. In \sec{sec:Review}
we review the setup of our manifestly gauge invariant
ERG. \Sec{sec:Methodology} is devoted to the
methodology for computing the expectation
of gauge invariant operators. The basic
idea, for which little prior knowledge
is required, is detailed in the short \sec{sec:basics}.
In the remainder of \sec{sec:Methodology}, the
machinery for performing calculations in
perturbation theory is developed. This section
concludes with a fantastically compact diagrammatic 
expression for the perturbative corrections
to the expectation value of any gauge invariant
operator.
In \sec{sec:WL}, we specialize to the computation
of the expectation values of (renormalized) Wilson loops.
After covering some general features in \sec{sec:WL-Gen}, 
in \sec{sec:WL-TxL}
we compute the $\Or (g^2)$
correction to the Euclidean, rectangular Wilson loop
with sides $T \gg L$ and recover the standard result.
We conclude in~\sec{sec:Conc}.

\section{Review}
\label{sec:Review}

\subsection{Elements of $SU(N|N)$ Gauge Theory}
\label{sec:elements}

Throughout this paper, we work in Euclidean dimension, D.
We regularize $SU(N)$ Yang-Mills, carried by
the physical gauge field $A^1_\mu$, by embedding
it in spontaneously broken $SU(N|N)$ Yang-Mills,
which is itself regularized by covariant higher derivatives~\cite{SU(N|N)}.
The massive gauge fields arising
from the spontaneous symmetry breaking play the
role of gauge invariant Pauli-Villars (PV) fields, furnishing the
necessary extra regularization to supplement the covariant
higher derivatives. In order
to unambiguously define contributions which are finite
only by virtue of the PV regularization, a preregulator must
be used in $D=4$~\cite{SU(N|N)}. We will use
dimensional regularization, emphasising that
this makes sense
non-perturbatively, since 
it is not being used to renormalize the theory, but rather
as a prescription for discarding surface terms in loop
integrals~\cite{SU(N|N)}.

The supergauge invariant Wilsonian effective action
has an expansion in terms of supertraces and
products of supertraces~\cite{aprop}:
\begin{widetext}
\begin{eqnarray}
	S 	& =	& \sum_{n=1}^\infty \frac{1}{s_n} \Int{x_1} \!\! \cdots \volume{x_n} 
				S^{X_1 \cdots X_n}_{\,a_1 \, \cdots \, a_n}(x_1, \cdots, x_n) \str X_1^{a_1}(x_1) \cdots X_n^{a_n} (x_n)
	\nonumber \\
		& +	& \frac{1}{2!} \sum_{m,n=0}^{\infty} \frac{1}{s_n s_m} \Int{x_1} \!\!\cdots \volume{x_n}  \volume{y_1} \!\! \cdots \volume{y_m}
				S^{X_1 \cdots X_n, Y_1 \cdots Y_m}_{\, a_1 \, \cdots \, a_n \, , \, b_1\cdots \, b_m}(x_1, \cdots, x_n; y_1 \cdots y_m)
	\nonumber \\
		&	&	
			\qquad \times \str X_1^{a_1}(x_1) \cdots X_n^{a_n} (x_n) \, \str Y_1^{b_1}(y_1) \cdots Y_m^{b_m} (y_m)
		\qquad +  \ldots 
\label{eq:ActionExpansion}
\end{eqnarray}
\end{widetext}
where the $X^{a_i}_i$ and $Y^{b_j}_j$ are embeddings of broken phase
fields into supermatrices.
 We take only one cyclic ordering for the 
lists $X_1 \cdots X_n$, $Y_1 \cdots Y_m$ in the sums over $n,m$.
If any term is invariant under some nontrivial cyclic
permutations of its arguments, then $s_n$ ($s_m$) is the order of the
cyclic subgroup, otherwise $s_n = 1$ ($s_m = 1$).

The momentum space vertices are written
\beas
	\lefteqn{S^{X_1 \cdots X_n}_{\,a_1 \, \cdots \, a_n}(p_1, \cdots, p_n) \left(2\pi \right)^D \delta \left(\sum_{i=1}^n p_i \right)} \\
	& &
	=
	\Int{x_1} \!\! \cdots \volume{x_n} e^{-i \sum_i x_i \cdot p_i} S^{X_1 \cdots X_n}_{\,a_1 \, \cdots \, a_n}(x_1, \cdots, x_n),
\eeas
where all momenta are taken to point into the vertex. We will employ the shorthand
\[
	S^{X_1 X_2}_{\, a_1 \; a_2}(p) \equiv S^{X_1 X_2}_{\, a_1 \; a_2}(p,-p).
\]

In addition to the coupling, $g$, of the physical gauge
field, there is a second dimensionless coupling, $g_2$,
associated with one of the unphysical regulator 
fields, $A^2_\mu$~\cite{aprop,Thesis,mgierg1,mgierg2}.
For convenience, we work not with $g_2$ directly
but with
\be
	\alpha \equiv g^2_2/g^2.
\label{eq:alpha-defn}
\ee
The coupling $g$ (similarly $\alpha$) 
is defined through its
renormalization condition:
\be
\label{defg}
	S[A^1] = {1\over2g^2}\,\tr\!\int\!\!d^D\!x\,
	\left(F^1_{\mu\nu}\right)^2+\cdots,
\ee
where 
the ellipses stand for higher dimension operators and the
ignored vacuum energy. \Eqn{defg}
constrains the classical two-point  vertex of
the physical field,  $S_{0 \mu \ \; \nu}^{\ A^1 A^1}(p) \equiv S_{0 \mu  \nu}^{\ 1 \, 1}(p)$, as follows:
\be
\label{eq:S_0-11}
	S_{0 \mu  \nu}^{\ 1 \, 1}(p) =  2 (p^2 \delta_{\mu\nu} - p_\mu p_\nu) + \Op{4}
	\equiv 2\Box_{\mu \nu}(p) + \Op{4}.
\ee

\subsection{Diagrammatics}

In this section, we introduce and
describe the diagrammatics necessary
for this paper. For a comprehensive description
of the diagrammatics see~\cite{Thesis,mgierg1}.

\subsubsection{Diagrammatics for the Action}

The \emph{vertex coefficient functions} belonging to the
action~\eq{eq:ActionExpansion} have a simple diagrammatic
representation:
\be
	\cd{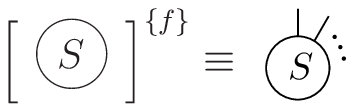}
\label{eq:WEA-VCF}
\ee
represents all vertex coefficient functions corresponding
to all cyclically independent orderings of the
set of broken phase fields, $\{f\}$, distributed
over all possible supertrace structures. For example,
\be
\label{eq:Vertex-A1A1}
	\dec{
		\SVertex
	}{A^1A^1}
\ee
represents the coefficient functions $S^{A^1 A^1}$
which, from~\eq{eq:ActionExpansion}, is associated
with the (super)trace structure $\tr A^1 A^1$. This
diagram would also represent the coefficient
function $S^{A^1, A^1}$, were it not for the fact that
this does not exist, on account of $\tr A^1 =0$.

\subsubsection{Diagrammatics for the Exact Flow Equation}

The diagrammatic representation of the flow
equation is shown in \fig{fig:Flow}~\cite{Thesis,mgierg1}.
\bcf[h]
	\[
	\cd{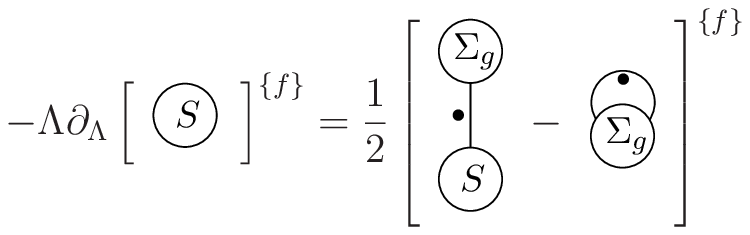}
	\]
\caption{The diagrammatic form of the flow equation.}
\label{fig:Flow}
\ecf

The \lhs\ just depicts the flow of all cyclically independent
Wilsonian effective action
vertex coefficient functions.
The objects on the \rhs\ of \fig{fig:Flow}
have two different types of component. The lobes
represent vertices of action functionals,
where $\Sigma_g \equiv g^2S - 2 \hat{S}$,
$\hS$ being the seed action~\cite{Thesis,mgierg1,mgierg2,scalar1,scalar2,aprop,giqed}:
a functional which respects the same symmetries as the Wilsonian effective 
action, $S$, and has the same structure. 
Physically, the seed action can be thought of as (partially)
parameterizing a general Kadanoff blocking in the 
continuum~\cite{jose,mgierg1}.

The object attaching
to the various lobes, \DummyKernel,  is
the sum over vertices of the covariantized 
ERG kernels~\cite{ymi,aprop}
and, like the action vertices, can be 
decorated by fields belonging to $\{f\}$.
The fields of the action vertex (vertices) to
which the vertices of the kernels attach
act as labels for the ERG kernels.
We loosely refer to both individual and summed over 
vertices of the kernels simply as a kernel.

The rule for decorating the diagrams on the \rhs\
is simple: the set of fields, $\{f\}$, are distributed in 
all independent ways between the 
component objects of each diagram.

Following~\cite{ym,ymi,ymii,aprop,Thesis,mgierg1,scalar2},
it is technically conventient to  use the freedom inherent in 
$\hat{S}$ by choosing the two-point, classical
seed action vertices equal to the corresponding Wilsonian effective
action vertices. The effect of this is that the kernels,
integrated \wrt\ $\ln \Lambda$ (at constant $\alpha$),
turn out to the the inverses 
of the classical, two-point vertices in the transverse
space.
For example, in the $A^1$-sector we find that
\be
\label{eq:EP-A1}
	S^{\ 1 \, 1}_{0 \mu \alpha} (p) \Delta^{1  \,1}_{\alpha \nu} (p) = \delta_{\mu\nu} - \frac{p_\mu p_\nu}{p^2},
\ee
where $\Delta^{1 1}$ in the integrated $A^1$ sector kernel.
It is apparent that $\Delta^{1 1}$ is the inverse of the 
corresponding classical, two-point
vertex up to a remainder term which, since it is forced to be
there as a consequence of the manifest gauge invariance, we
call a `gauge remainder'. In recognition
of the similarities of the integrated kernels to
propagators, in both form and diagrammatic role, we refer to them
as effective propagators~\cite{aprop}. However, we emphasise that
at no point is gauge fixing required in their definition
and that our diagrams do not correspond, in any way,
to conventional Feynman diagrams. \Eqn{eq:EP-A1} can be 
diagrammtically generalized to hold in all sectors:
\be
	\resizebox{8.5cm}{!}{\includegraphics{EPR.epsi}}
\label{eq:EPReln-A}
\ee
We have attached the effective propagator,
denoted by a solid line, 
to an arbitrary structure since it only
ever appears as an internal line.
The field labelled by $M$ can be any of the broken phase
fields. The object $\GR \!\! \equiv \; \GRkpr \!\!\! \GRk$ 
is a gauge remainder.
The individual components of
$\GRkpr \!\! \GRk$ will often be loosely
referred to as gauge remainders; where it is necessary to
 unambiguously refer to the composite structure, we will use
the terminology `full gauge remainder'.
\Eqn{eq:EPReln-A} is referred
to as the effective propagator relation.
From~\eqs{eq:S_0-11}{eq:EP-A1}, it follows
that
\be
\label{eq:EP-leading}
	\Delta^{11}_{\rho \sigma}(p) = \frac{\delta_{\rho \sigma}}{2 p^2}
	(1 + \Or (p^2/\Lambda^2)),
\ee
which we will use later.

Embedded within the diagrammatic rules is a prescription for evaluating the
group theory factors. 
Suppose that we wish to focus on the flow of a particular
vertex coefficient function which, necessarily, has a unique
supertrace structure.

On the \rhs\ of the flow equation, we must
focus on the components of each diagram
with precisely the same
supertrace structure as the \lhs,
noting that the kernel, like the vertices,
has multi-supertrace contributions (for more
details see~\cite{Thesis,mgierg1}).
In this more explicit diagrammatic picture,
the kernel is to be considered a double
sided object.
Thus, whilst the dumbbell like term of \fig{fig:Flow}
has at least one associated supertrace, the next diagram
has  at least two, on a account of the loop
(this is strictly true only in the
case that kernel attaches to fields on the same
supertrace). If a closed
circuit formed by a kernel is devoid
of fields then
it contributes 
a factor of $\pm N$, depending on
the flavours of the fields to which the kernel forming
the loop attaches. This is most easily appreciated by
defining the projectors
\[
	\sigma_{\pm} \equiv \hf ( \one \pm \sigma)
\]
and noting that $\str \sigma_\pm = \pm N$. 
In the counterclockwise sense, either
a $\sigma_+$ or $\sigma_-$, as appropriate,
can always be inserted for free after any
of the broken phase fields.

The above prescription for evaluating the
group theory factors receives $1/N$ corrections in 
the $A^1$ and $A^2$ sectors. If a kernel
attaches to an $A^1$ or $A^2$, it comprises a direct
attachment and an indirect attachment.
In the former case, one supertrace associated
with some vertex coefficient function  is `broken
open' by an end of a kernel: the fields on
this supertrace and the single supertrace component of the
kernel are on the same circuit.
In the latter case, the kernel does not break anything open
and so the two sides of the kernel pinch together
at the end associated with the indirect attachment.
This is illustrated
in \fig{fig:Attach}; for more detail, see~\cite{Thesis,mgierg1}.
\bcf
	\[
	\resizebox{8.5cm}{!}{\cd{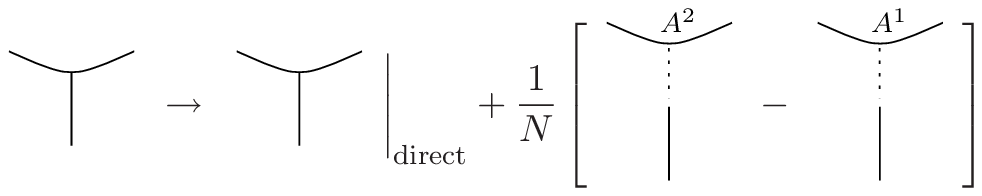}}
	\]
\caption{The $1/N$ corrections to the group theory factors.}
\label{fig:Attach}
\ecf 

We can thus consider the diagram on the \lhs\ as having been unpackaged,
to give the terms on the \rhs. The dotted lines in the diagrams with indirect
attachments serve to remind us where the loose end of the kernel attaches
in the parent diagram.

As an example, which will be of use later,
consider the group theory
factor of the diagram on the \lhs\ of \fig{fig:GroupTheory},
where we suppose that the kernel forming the loop is in
the $A^1$ sector.
\bcf
	\[
	\resizebox{8.5cm}{!}{\cd{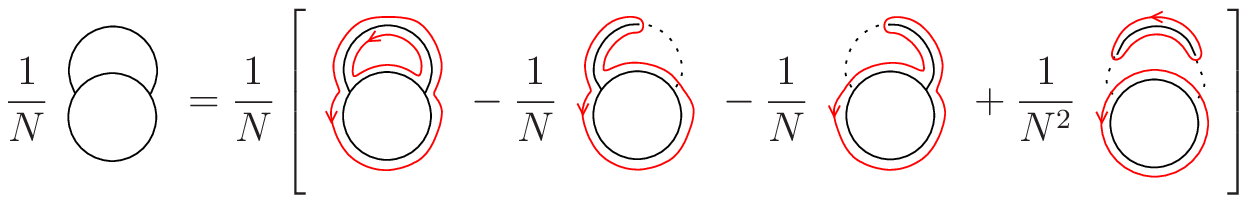}}
	\]
\caption{An example showing how to evaluate the group 
theory factor of a diagram in which the kernel is taken to
be in the $A^1$ sector.}
\label{fig:GroupTheory}
\ecf

On the \rhs, we have unpackaged the parent diagram and explicitly
indicated, in red, how many circuits each diagram has.
To evaluate the corresponding group theory factors,
we simply take each circuit
to contribute $\str \sigma_+$
($\sigma_+$ because we are taking the kernel to be in the $A^1$
sector). Therefore, the overall group theory factor is
\[
	\frac{1}{N} \left( N^2 - 2\frac{1}{N} N + \frac{1}{N^2} N^2 \right) 
	= \frac{1}{N} (N^2 - 1) = 2 C_2(F),
\]
where $C_2(F)$ is the quadratic Casimir operator for
the fundamental representation of $SU(N)$.

\section{Methodology}
\label{sec:Methodology}

\subsection{Basics}
\label{sec:basics}

In this section we describe the strategy for
computing the renormalized
expectation value of the
gauge invariant operator, $\GIO$. Denoting the
set of (dynamical) broken phase fields by $\fields$,
we aim to compute
\be
	\expectation{\GIO}_R = 
	\frac{1}{Z_0} \int_{\Lambda_0} \measure{\fields} \GIO_{\Lambda_0}[A^1] \, e^{-S_{\Lambda_0}[\fields]},
\label{eq:expectation}
\ee
where the subscript $R$ 
stands for renormalized and
$Z_0 = \int_{\Lambda_0} \measure{\fields} e^{-S_{\Lambda_0}[\fields]}$.
Notice that we have explicitly tagged the functional integral,  action
and $\GIO$ with $\Lambda_0$. This is to remind us that the expression is
defined at the bare scale, $\Lambda_0$. 
At this scale, the gauge invariant
operator is taken to be a functional of just the physical gauge field, $A^1$.
The limit 
$\Lambda_0 \rightarrow \infty$,
which essentially corresponds to the
continuum limit~\cite{TRM-elements},
is taken at the end of a calculation.

Introducing the source, $J$, we rewrite~\eq{eq:expectation} in the usual way:
\be
\label{eq:expectation-b}
	\expectation{\GIO}_R =  - \left. \pder{}{J} \ln Z_J \right|_{J=0},
\ee
where
\[
	Z_J = \int_{\Lambda_0} \measure{\fields} e^{-S_{\Lambda_0}[\fields] - J \GIO_{\Lambda_0}[A^1]}.
\]
The key step now is to integrate out
modes between the bare scale and the 
effective scale, $\Lambda$, to yield:
\be
\label{eq:ZJ-effective}
	Z_J = \int_{\Lambda} \measure{\fields} e^{-S_{\Lambda}[\fields] - \GIO_{\Lambda}[\fields, J]}.
\ee
Since both $S_\Lambda$ and $\GIO_\Lambda$ are gauge invariant, the division of terms
between these two functionals is of course arbitrary. For example,
for some other definition of $S_\Lambda$ and $\GIO_\Lambda$, we could have
written the argument of the exponential in~\eq{eq:ZJ-effective} as
$S'_{\Lambda}[\fields, J] - \GIO'_{\Lambda}[\fields, J]$ or even
just $S''_{\Lambda}[\fields, J]$. However, we choose to define
things such that $S_\Lambda$ is independent of $J$. Given that
the only dependence on $J$ at the bare scale is linear, it follows
from the flow equation that the dependence at the effective scale
has a Taylor expansion in $J$
and so we can write:
\be
\label{eq:GIO-eff}
	\GIO_{\Lambda}[\fields, J] = \sum_{i=1}^\infty J^i \GIO^i_\Lambda[\fields].
\ee
The real point now is that, from~\eq{eq:expectation-b},
we are pulling out the $\order{J}$ part, only, when
computing the expectation value. Therefore, it makes
sense to work at small $J$, in which case we can take
the effect of introducing the gauge invariant operator
at the bare scale as inducing an infinitesimal, linear
perturbation to the Wilsonian effective action
at the effective scale~\cite{ymi}:
\be
\label{eq:Action-Pert}
	S_\Lambda \rightarrow S_\Lambda + J \GIO_\Lambda + \order{J^2}
\ee
where, since we are henceforth only
interested in the $\order{J}$ part of~\eq{eq:GIO-eff}, 
we have dropped the superscript index of $\GIO^1_\Lambda$.

By performing the shift~\eq{eq:Action-Pert} in the flow
equation, we see that
the flow of the Wilsonian effective action is still given
as in \fig{fig:Flow} and the flow of $\GIO_\Lambda$ is given
in \fig{fig:Oflow}, where we define
\be
\label{eq:Pi}
	\Pi \equiv g^2 S - \hat{S},
\ee
take squares to represent vertices of $\GIO_\Lambda$
and have dropped the subscript $\Lambda$.
\bcf
	\[
	\cd{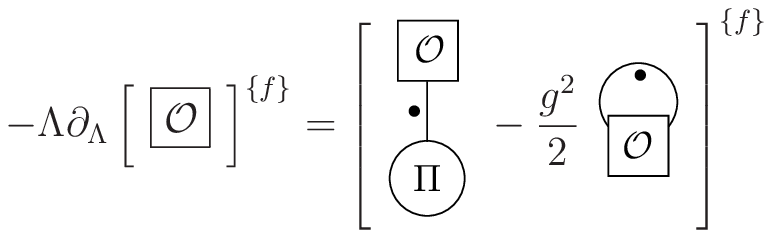}
	\]
\caption{The flow of $\GIO_\Lambda$.}
\label{fig:Oflow}
\ecf

Next, consider how $Z_J$ evolves as we integrate out
all the modes \ie\ as we take the limit $\Lambda \rightarrow 0$.
Let us start with the behaviour of the gauge invariant
operator, $\GIO$. Like the Wilsonian effective action, $\GIO$ has
an expansion in terms of fields. However, unlike in~\eq{eq:ActionExpansion},
it is crucial that we retain the field-independent part
(\ie\ the vacuum energy-like term). As we integrate
out modes, so this term receives quantum corrections. What
of the field dependent parts? Clearly, once we have integrated
out all modes, there cannot be any field dependent terms remaining
which are multiplied by a finite coefficient. There are two choices:
either the coefficients diverge, in which case 
$e^{-\GIO_{\Lambda \rightarrow 0}} \rightarrow 0$,
or each coefficient corresponding to a field dependent
term in the expansion of $\GIO$ vanishes. We assume that the latter
is the case. 

In the case of the Wilsonian effective action,
matters are simple. The structure of~\eq{eq:expectation}
ensures that, when computing 
$\expectation{\GIO}$, the factor 
of $e^{-S_{\Lambda \rightarrow 0}}$ in the numerator
is cancelled out by the $Z_0$ 
in the denominator.\footnote{
In fact, there are terms in the Wilsonian effective action which do diverge
as $\Lambda \rightarrow 0$~\cite{SU(N|N)}.
This is easy to see: in order
for the regularization scheme to work, the effective propagator
in the $A^1$ sector dies off, for $p^2/\Lambda^2 \gg 1$, at least
as fast as $(p^2/\Lambda^2)^{-r}$ (for $r>2$).
This means that the kinetic term must be modified
by a term which behaves as $(p^2/\Lambda^2)^{+r}$ for $p^2/\Lambda^2 \gg 1$.
Such terms in the effective action clearly diverge as $\Lambda \rightarrow 0$.
}

Therefore, from~\eq{eq:expectation-b}, \eq{eq:ZJ-effective}, \eq{eq:GIO-eff}
and~\eq{eq:Action-Pert} we deduce the beautifully simple equation
\be
\label{eq:GIO-zero}
	\expectation{\GIO}_R = \GIO_{\Lambda =0}.
\ee

To find $\expectation{\GIO}_R$, we can use~\fig{fig:Oflow}
to compute the flow of $\GIO_\Lambda$,
\fig{fig:Flow} to compute the flow of $S$ (which is buried in $\Pi$)
and thereby determine $\GIO_{\Lambda =0}$, in some approximation.
For the remainder of this
paper, we will work in the perturbative domain. 
Dropping the $\Lambda$, which we
now take to be implicit, we take the following weak
coupling expansions.
The Wilsonian effective action is given by
\be
	S = \sum_{i=0}^\infty g^{2(i-1)} S_i = \frac{1}{g^2}S_0 + S_1 + \cdots,
\label{eq:Weak-S}
\ee
where $S_0$ is the classical effective action and the $S_{i>0}$
the $i$th-loop corrections; $\GIO$ is given by
\be
\label{eq:GIO}
	\GIO = \sum_{i=0}^\infty g^{2(i-1)} \GIO_i.
\ee
The seed action has an expansion consistent with
the fact that $S$ appears in the flow equation multiplied
by an extra power of $g^2$, compared to $\hS$:
\be
	\hat{S} = \sum_{i=0}^\infty  g^{2i}\hat{S}_i.
\label{eq:Weak-hS}
\ee
Recalling~\eq{eq:alpha-defn} we have:
\bea
	\beta \equiv \flow g 		& = & \sum_{i=1}^\infty  g^{2i+1} \beta_i(\alpha)
\label{eq:beta}
\\[1ex]
	\gamma \equiv \flow \alpha 	& = & \sum_{i=1}^{\infty}  g^{2i} \gamma_i(\alpha).
\label{eq:gamma}
\eea

To obtain the weak coupling flow equation for $\GIO$
we substitute~\eq{eq:Weak-S}--\eq{eq:gamma}
into~\fig{fig:Oflow}, 
but do not preclude the possibility
that, in addition to $g$ and $\alpha$,
$\GIO_\Lambda$ also depends on
the dimensionless, running 
couplings $\alpha^{i>1}$
(we identify $\alpha^1$ with $\alpha$).
This anticipates our treatment of Wilson
loops in \sec{sec:WL}.
The
weak coupling flow equations for $\GIO$
are shown in \fig{fig:WeakCoupling-O},
where $\bigdot{X} \equiv -\flowConstAl X$,
$n_r \equiv n-r$, $n_- \equiv n-1$
and $\Pi_r \equiv S_r - \hS_r$.
\bcf
	\[
	\resizebox{8.5cm}{!}{\cd{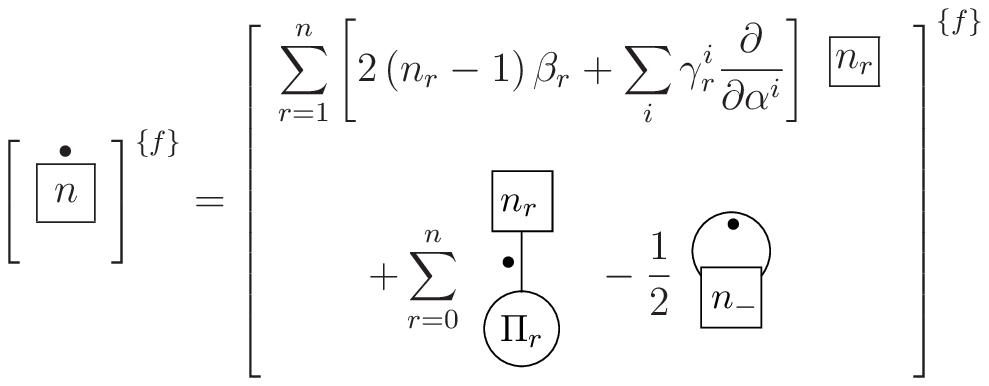}}
	\]
\caption{The weak coupling flow equations for $\GIO$.}
\label{fig:WeakCoupling-O}
\ecf

\subsection{Additional Notation}

In the diagrammatic flow equation---be it the exact form
or the perturbative expansion---we have considered
decoration by the set of fields $\{f\}$. However, only
on the \rhs\ of~\eq{eq:WEA-VCF} have we actually converted the
fields $\{f\}$ into explicit decorations.
Before such decoration, we
consider $\{f\}$ to be implicit, or unrealized, 
decorations~\cite{Thesis,Primer}.
Just as it is useful to consider fields as implicit decorations,
so too is it useful to construct rules
for decoration with  implicit effective propagators
and instances of the gauge remainder component $\GRkpr$. 

\subsubsection{Gauge Remainders}

Instances of $\GRkpr$
arise from diagrams in which the effective
propagator relation~\eq{eq:EPReln-A} has been
applied, generating a full gauge remainder.
The $\GRk$ part of the gauge remainder
can be processed, using the Ward identities~\cite{ymi,aprop,mgierg1,Thesis},
leaving behind a $\GRkpr$.
If one of the vertices generated by
the $\GRk$ is a classical, two-point vertex,
then in the case where this vertex attaches to an effective
propagator, a further full gauge remainder is generated.
Processing this gauge remainder using the Ward identities
allows us to iteratively generate 
structures containing an arbitrary number, $m$, of $\GRkpr$s.
We denote $m$ implicit instances of 
$\GRkpr$ by
\[
	\dec{ \
	}{\GRkpr^m \cdots},
\]
where the square brackets could enclose some
diagrammatic structure, but need not. The ellipsis
represents any additional implicit decorations,
so long as they are not further instances of $\GRkpr$.
The superscript notation $\GRkpr^m$ simply tells
us that there are $m$ instances of $\GRkpr$.

For the purposes of this paper, we do not require the
rules for turning gauge remainders appearing as 
implicit decorations into explicit structures. The
details can be found in~\cite{mgiuc}.

\subsubsection{Effective Propagators}

The rule for explicit decoration with implicit
effective propagators is as follows.
If we wish to join two objects
(say two vertices) together with $j'$ out of a total
of $j$ effective propagators, then there are
$\nCr{j}{j'} 2^{j'}$ ways to do this. 
Intuitively, the first factor captures the 
notion that,
so long as they are implicit decorations, the effective
propagators are indistinguishable. The factor of
$2^{j'}$ allows for the fact that we can interchange
the two ends of an effective propagator. If these
effective propagators were instead used to form $j'$
loops on a single vertex, then the factor of
$2^{j'}$ would disappear, since the vertices
are defined such that all cyclically independent
arrangement of their decorative fields are summed
over.

\subsubsection{Vertices}

When analysing the perturbative flow of $\GIO$,
we will find that vertices (of the Wilsonian effective
action and $\GIO$) always occur in a very particular way.
To introduce compact notation for this, we start by 
introducing a set of vertex arguments, $v^j$,
where the upper roman index acts as a label. Thus, the $v^j$
are integers, denoting the loop orders of some set of vertices.
In the case that a vertex argument labels a Wilsonian
effective action vertex, we define the reduction of $v^j$,
$v^{j;R}$, such that a reduced vertex does not have a classical,
two-point component.

Next, we define
\beas
	v^{j,j_+} & \equiv & v^j - v^{j+1},
\\
	v^{j,j_+;R} & \equiv & v^{j;R} - v^{j+1;R}.
\eeas
and use this notation to construct
\be
\label{eq:VertexTower}
	\left[ \OTower \right]
	\equiv
	\prod_{i=0}^{j-1} \sum_{v^{i_+} = 0}^{v^i}
	\left[
		\sco[2]{\TopO}{\Vertex{v^{i,i_+;R}}}
	\right],
\ee
where $n_s$ gives the value of $v^0$, which is the only vertex argument
not summed over on the \rhs. 
Notice that the sum over all vertex arguments is trivially $n_s$:
\be
\label{eq:VertexSum}
	\sum_{i=0}^{j-1} v^{i,i_+} + v^{j} = \sum_{i=0}^{j-1} \left( v^i - v^{i+1} \right) + v^{j} = v^0 = n_s.
\ee

The structure defined by~\eq{eq:VertexTower}
possesses a single vertex belonging to $\GIO$
and $j$ (reduced) Wilsonian effective action
vertices. This allows us to usefully
define~\eq{eq:VertexTower} for $j=0$:
\[
	\left[
		\sco[2]{\TopO}{\SumVertex}
	\right]_{j=0} \equiv
	\OVertex{n_s}.
\]

\subsection{The Diagrammatic Function, $\Q_n$}

We introduce the functions
\be
\label{eq:Q}
\Q_n \equiv \OVertex{\; n \;}
	{\small -2 \sum_{s=1}^n  \sum_{m=0}^{2s-1} \sum_{j=0}^{n+s-m-1}
	\frac{\norm_{j+s,j}}{m!}}
	\dec{\!\!
		\sco[1]{\TopO}{\SumVertex} \rule{0cm}{1.1cm}\!\!
	}{\tiny \Delta^{j+s}\GRkpr^m}
\ee
\be
\label{eq:Qbar}
	\bar{\Q}_n  \equiv  \OVertex{\; n \;} - \Q_n
\ee
where, for the non-negative integers $a$ and $b$,
we define
\be
\label{eq:norm}
	\norm_{a,b} = \frac{(-1)^{b+1}}{a!b!} \left(\frac{1}{2}\right)^{a+1}.
\ee
In the case that either $a$ or $b$ are negative, $\norm_{a,b}$
is null. 
As part of the definition of $\bar{\Q}$, we insist
that, upon explicit decoration, all fully fleshed out diagrams
must be connected.

There is a simple, intuitive explanation for the relationship
between the total number of vertices, the number of effective propagators
and the sum over the vertex arguments. This is most simply put by
taking $m=0$ (the following argument is easily generalized).
From~\eq{eq:VertexSum}, we know that the sum of the vertex arguments is $n-s$.
Now, given
$j+1$ vertices, exactly $j$ effective propagators
are required to create a connected diagram. This leaves
over $s$ effective propagators, each of which must create
a loop. Therefore, the loop order of the diagram is $n-s+s = n$,
as must be the case.

The maximum values of the sums over $m$ and $j$
follow from the constraint
that all fully fleshed out diagrams are connected~\cite{mgiuc}.
The maximum
value of $s$ clearly follows from the requirement
that the loop order of the diagram is $\geq 0$. 
The minimum value of $s$ ensures that, in $\Q$,
we do not double count the contribution $\OVertex{n}$.

\subsection{Expectation Values in Perturbation Theory}

The key to computing expectation values is
to consider the flow of $\Q_n$.
It can be shown~\cite{mgiuc}, by using the flow equation,
that this yields\footnote{This result, though intuitive,
is far from straightforward to derive, afresh. However,
the more difficult case of deriving similar diagrammatic expressions
for the perturbative $\beta$-function coefficients is
comprehensively illustrated in~\cite{mgiuc}. Given
this derivation, \eq{eq:flow-Q} follows, essentially trivially.}:
\be
\label{eq:flow-Q}
	\flow \Q_n	+
	2\sum_{n'=1}^{n} (n -n'-1)\beta_{n'} \Q_{n-n'} =0,
\ee
from which it follows that
\[
	\totalflow \sum_{n=0}^\infty
	\left[
		g^{2(n-1)} \Q_n
	\right]
	= 0.
\]
Integrating between $\Lambda = \mu$ and $\Lambda = \Lambda_0$ gives
\bea
	\lefteqn{\sum_{n=0}^\infty
	\left[
		g^{2(n-1)} 
		\left(
			\OVertex{\; n \;} 
			- \bar{\Q}_n
		\right)
	\right]_{\Lambda=\mu}}
	\nonumber \\
\label{eq:Integrate}
	& &=
	\sum_{n=0}^\infty
	\left[
		g^{2(n-1)} 
		\left(
			\OVertex{\; n \;} 
			- \bar{\Q}_n
		\right)
	\right]_{\Lambda=\Lambda_0},
\eea
where we have used~\eq{eq:Qbar} and we aim to
take the limits $\mu \rightarrow 0$ and $\Lambda_0 \rightarrow \infty$.

The crucial point to recognize now is that
(in perturbation theory, at any rate)
\be
\label{eq:vanish}
	\lim_{\Lambda \rightarrow 0}
	\left[
		g^{2(n-1)} \bar{\Q}_n
	\right] = 0.
\ee
We can argue this as follows.
Consider $\bar{\Q}_n$,
which is both UV and IR finite,
in the limit of small $\Lambda$.
At the level of the diagrammatic components
out of which $\bar{\Q}_n$ is built, 
all contributions for which the 
loop momenta $k^i \gg \Lambda$
are suppressed by the UV regularization.
(We might worry that this suppression
does not occur in diagrams possessing
classical vertices
which diverge in this limit. However,
these divergences are always overcompensated.)
Thus, in the limit $\Lambda \rightarrow 0$,
the loop integrals have no support and
$\bar{\Q}_n$ vanishes.

To complete the argument, all
that remains to be done is to
show that the behaviour of $\lim_{\Lambda \rightarrow 0} g^{2(n-1)}(\Lambda)$
is sufficiently good. 
It should be emphasised
that we are applying this limit to 
quantities computed in perturbation theory. 
Introducing the arbitrary scale, $M$,
we can write
\[
	g^{2}(\Lambda) =  \sum_{i=1}^{\infty} g^{2i}(M) a_i(M/\Lambda).
\]
Differentiating both sides
\wrt\ $M$ yields the set of relationships:
\[
	0 = 2 \sum_{j=1}^{n-1} (n-j) \beta_j a_{n-j}(M/\Lambda) 
	+ \der{a_n(M/\Lambda)}{\ln M/\Lambda}.
\]
Given that $a_1 =1$, it follows 
that every $a_i$ must be a
function of $\ln M/ \Lambda$.
Therefore, in the $\Lambda \rightarrow 0$
limit, $g(\Lambda)$ diverges, at worst, as
powers of $\ln \Lambda$. This growth
is slower than the rate at which the
UV regularization kills $\bar{\Q}_n$ in
the limit that $\Lambda \rightarrow 0$~\cite{SU(N|N)}
and so we have demonstrated~\eq{eq:vanish}.

From~\eqs{eq:Integrate}{eq:vanish}, we arrive at
the central result of our perturbative treatment:
\bea
	\lefteqn{\sum_{n=0}^{n'}
	\left[
		g^{2(n-1)} 	\OVertex{\; n \;} 
	\right]_{\Lambda=0}} \nonumber \\
\label{eq:Central}
	& & =
	\sum_{n=0}^{n'}
	\left[
		g^{2(n-1)} 
		\left(
			\OVertex{\; n \;} 
			- \bar{\Q}_n
		\right)
	\right]_{\Lambda=\Lambda_0} + \Or (g^{2n'}). \qquad \qquad
\eea
Notice that we have replaced the upper limits
of the sums over $n$ with the finite $n'$.
By taking $n'$ to infinity, \eq{eq:Central} becomes
exact. However, the form given above is suitable for
the order-by-order computation of corrections
to $\expectation{\GIO}_R$.

\section{Wilson Loops}
\label{sec:WL}

In this section, we will illustrate~\eq{eq:Central}
by using it to compute perturbative corrections
to Wilson loops. Before taking the explicit case
of the rectangular Wilson loop with sides $T \gg L$,
we discuss some general features of Wilson loop
calculations, within our framework.

\subsection{General Considerations}
\label{sec:WL-Gen}

For some closed path, $\Gamma$, the
path ordered phase factor, \aka\ the Wilson
loop, is defined to be
\[
	\WL{\Loop} = 
	\frac{1}{N} \tr P \exp 
	\left[
		i \oint_\Gamma dx_\mu A^1_\mu(x)		
	\right].
\]
It is well known~\cite{Poly,D+V,B+N+S} that 
the expectation value of this object,
\be
\label{eq:EWL}
	\EWL{\Loop} = \expectation{\WL{\Loop}},
\ee
is divergent even after renormalization of
the coupling and, in the case of a gauge
fixed formulation, field strength renormalization.
In our manifestly gauge invariant
formulation, where the gauge field does not
suffer from field strength renormalization, 
\eq{eq:EWL} is defined such that the renormalization
of the coupling has been done.

The remaining divergences have two sources.
For smooth, simple loops, there
is a divergence $e^{-\kappa \Lambda_0 \length{\Loop}}$,
where $\kappa$ is a dimensionless parameter and 
$\length{\Loop}$
is the length of $\Gamma$.
The linearly divergent
$K \equiv \kappa \Lambda_0$ can
be interpreted as a mass divergence.
The other divergences
come from any (finite)
number of cusps and intersections,
parameterised by the angles $\cusps^i$ and $\ints^i$, respectively.
The renormalized expectation value of the
Wilson loop with cusps but no intersections is
defined to be~\cite{B+N+S}
\[
	\REWL{\Loop_{\cusps^i}} = Z(\cusps^i) e^{-\overline{m}_0 \Lambda_0 \length{\Loop}} \EWL{\Loop_{\cusps^i}},
\]
where we have used powers of $\Lambda_0$ to
replace the bare mass, $m_0$, with a dimensionless
parameter, $\overline{m}_0$.
The renormalized mass, $m$, is 
\[
	m = K + m_0
\]
and the multiplicative renormalization constant factorizes:
\[
	Z(\cusps^i) = Z(\cusps^1)  Z(\cusps^2) \cdots.  
\]
(In the case that $\Loop$ includes intersections, $\REWL{\Loop}$
no longer renormalizes by itself, and must be considered
together with expectation values of a family
of other loop functions.)


$Z(\cusps)$ and $\overline{m}$ have the following expansions:
\beas
	Z(\cusps,g) & = & 1 + \sum_{i=1}^{\infty} g^{2i} Z_i(\cusps)
\\
	\overline{m}(g)		& =	& \sum_{i=1}^{\infty} g^{2i} \overline{m}_{i}.
\eeas

With these points in mind, we 
identify the boundary value of our
gauge invariant operator 
with 
\be
\label{eq:WL-def}
	\GIO_{\Lambda_0} = 
	\frac{1}{N} \tr P \exp 
	\left[
		i \oint_\Gamma dx_\mu A^1_\mu(x)		
	\right] e^{-\overline{m}_0 \Lambda_0 \length{\Loop}} Z(\cusps^i).
\ee

We can use the fact that $\GIO_{\Lambda_0}$ 
does not possess an $\order{1/g^2}$ component to
simplify the following analysis. To this end, consider
the classical flow of $\GIO$:
\[
	\cd{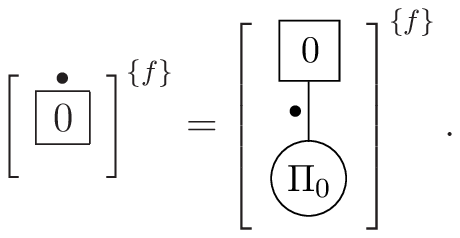}
\]
Now, for the \rhs\ not to vanish, the $\Pi_0$ vertex
must be decorated by at least two fields belonging to
$\{f\}$. This is because both seed action and Wilsonian
effective action one-point vertices vanish at tree 
level~\cite{aprop,mgierg1} and $\Pi_{0 R\, S}^{\ XX}(k) = 0$
due to our choice to set the seed action, two-point, classical
vertices equal to their Wilsonian effective action counter
parts.

From this it follows that
\[
	\stackrel{\bullet}{\OVertex{\; 0 \; }} = 0, \qquad
	\dec{
		\stackrel{\bullet}{\OVertex{\; 0 \; }}
	}{X} = 0,
\]
where $X$ is any field.
Integrating up and using the fact
that all classical vertices vanish at the boundary 
(see~\eqs{eq:GIO}{eq:WL-def}) we find that
\[
	\OVertex{\; 0 \; } = 0, \qquad
	\dec{
		\OVertex{\; 0 \; }
	}{X} = 0.
\]
But, these relationships, together with the
vanishing of $\Pi_{0 R\,S}^{\ XX}(k)$ 
and the boundary condition
imply that
\[
	\dec{
		\OVertex{\; 0 \; }
	}{XX} = 0.
\]
Iterating this argument, it is clear that, in fact,
\[
	\dec{
		\OVertex{\; 0 \; }
	}{\{f\}} = 0.
\]
Given that the $\GIO$ vertex of $\bar{\Q}$
must, therefore, have an argument of at least one,
this allows us to reduce the maximum
value of $j$ by unity~\cite{mgiuc}.

In a similar fashion, we can demonstrate that
\[
	\dec{
		\OVertex{\; 1 \; }
	}{X} = 0.
\]

With these points in mind, let us apply~\eq{eq:Central}
for $n'=1$.
Using the boundary condition, we
obtain the expected (trivial) result that
\be
\label{eq:WL-1}
	\left[ \OVertex{\; 1\;} \right]_{\Lambda=0} = 1.
\ee
[Note that~\eq{eq:WL-1} is in fact exact, not requiring
supplementation at $\Or (g^2)$. This follows
because, in the weak coupling expansion, the
vertex $\OVertex{\,1\,}$ multiplies $g^0$.]
At the next order we find
\be
\label{eq:WL-2}
\left[ g^2 \OVertex{\; 2\;} \right]_{\Lambda=0}
	=
	\left[
		g^2
		\left(
			\OVertex{\; 2 \;} 
			-2 \norm_{1,0} 
			\begin{array}{c}\begin{picture}(0,0)%
\includegraphics{pstex/0Padlock-1.pstex}%
\end{picture}%
\setlength{\unitlength}{3947sp}%
\begingroup\makeatletter\ifx\SetFigFont\undefined%
\gdef\SetFigFont#1#2#3#4#5{%
  \reset@font\fontsize{#1}{#2pt}%
  \fontfamily{#3}\fontseries{#4}\fontshape{#5}%
  \selectfont}%
\fi\endgroup%
\begin{picture}(391,576)(1801,-923)
\put(1946,-827){\makebox(0,0)[lb]{\smash{{\SetFigFont{11}{13.2}{\rmdefault}{\mddefault}{\updefault}{\color[rgb]{0,0,0}$1$}%
}}}}
\end{picture}%
 \end{array}
		\right)
	\right]_{\Lambda=\Lambda_0} + \Or (g^4).
\ee
Equation~\eq{eq:WL-2} gives the first
non-trivial correction to the renormalized
Wilson loop
parameterized by a contour with an arbitrary
(finite) number of cusps (generalization to include
intersections is straightforward, as indicated earlier). 
To evaluate~\eq{eq:WL-2} we feed in the boundary
condition~\eq{eq:WL-def}. The
first term on the \rhs\ possesses precisely
those contributions necessary to cancel
the divergences in the second term. With
these divergences cancelled, we can safely
take the continuum limit, $\Lambda_0 \rightarrow \infty$.

\subsection{The Rectangular Wilson Loop with sides $T \gg L$}
\label{sec:WL-TxL}

To illustrate the application of~\eq{eq:WL-2} in a way
which will allow us to compare directly
with known results, we must compute a quantity
which is independent of the renormalization prescription.
To this end, we focus on the 
rectangular Wilson loop, $\bar{\Loop}$, with sides $T$ and $L$,
in the  limit where $T \gg L$. 
The leading order contribution in this limit
is universal, being directly related
to the lowest order Coulomb potential of the
physical $SU(N)$ Yang-Mills theory.

At the boundary, the expression for the first term on
the \rhs\ of~\eq{eq:WL-2} follows, directly, from~\eq{eq:WL-def} 
upon expanding the exponentials and identifying the $\Or (g^2)$,
field independent contribution.
For the second term we must work a little harder,
since we need to relate
the two-point vertex to the boundary condition.
To do this, we expand
the exponential of~\eq{eq:WL-def} and focus on
the coefficient of $\tr A^1_\mu A^1_\nu$ at $\Or (g^0)$:
\begin{widetext}
\[
	-\frac{1}{2N} \oint_{\bar{\Loop}} dx_\mu \oint_{\bar{\Loop}} dy_\nu
	=  -\frac{1}{2N} 
		\Int{x} \ODInt{t} \der{x_\mu(t)}{t}\delta(x-x(t)) 
		\Int{y} \ODInt{s} \der{y_\nu(s)}{s}\delta(y-y(s)).
\]
The recasting on the \rhs\ allows us to directly compare
this expression with the field expansion of $\GIO$,
given by the analogue of~\eq{eq:ActionExpansion} with $S$ replaced by $\GIO$.
Therefore,
\[
	\cd{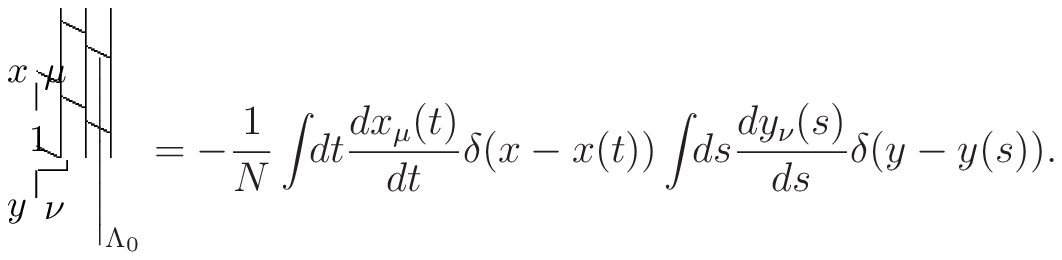}
\]
The other components of the second diagram on
the \rhs\ of~\eq{eq:WL-2} are the effective
propagator, 
$\Delta^{1\,1}_{\mu\nu}(x,y)$\footnote{
The fields must
be in the $A^1$ sector at the bare scale,
since this is the only sector in which the vertex to 
which the effective propagator attaches has support.
}, the group theory factor (which 
can be evaluated according to~\fig{fig:GroupTheory}) and an
integral over the undetermined coordinates, $x$ and $y$.

Using~\eq{eq:norm}, equation~\eq{eq:WL-2} becomes
\be
\label{eq:WL1-nearly}
	\lim_{T/L \rightarrow \infty} W^{(2)}_R(\bar{\Loop})= 
	\lim_{T/L \rightarrow \infty} \lim_{\Lambda_0 \rightarrow \infty}
	\left[
	g^2
	\left(
		\OVertex{\; 2 \;}
		-
		\frac{N^2-1}{2N} \oint_{\bar{\Loop}} dx_\mu \oint_{\bar{\Loop}} dx_\nu \Delta^{1\,1}_{\mu\nu}(x,y,\Lambda_0)
	\right)
	\right],
\ee
\end{widetext}
where we have changed notation slightly to make the path
dependence of the \lhs\ explicit.
Since we are taking the $T/L \rightarrow \infty$ limit,
we do not need to be too precise about our renormalization
prescription: the associated finite terms are sub-leading
and so we have:
\bea
\lefteqn{
	\lim_{T/L \rightarrow \infty} W^{(2)}_R(\bar{\Loop})= 
	\lim_{T/L \rightarrow \infty}
	\left[
	\rule{0cm}{0.5cm}
		- g^2 C_2(F)
	\right.} \nonumber \\
\label{eq:WL1-nearlyb}
	&& \qquad \qquad \times
	\left.
	\rule{0cm}{0.5cm}
		\left. 
			\oint_{\bar{\Loop}} dx_\mu 
			\oint_{\bar{\Loop}} dx_\nu \Delta^{1\,1}_{\mu\nu}(x,y) 
		\right|_{\mathrm{finite}}
	\right].
\eea
Writing
\[
	\Delta^{1\,1}_{\mu\nu}(x,y) = \IntB{p} e^{ip \cdot(x-y)} \Delta^{1\,1}_{\mu\nu}(p^2/\Lambda^2)
\]
and recalling~\eq{eq:EP-leading}, it is clear that 
our expression corresponds to the usual one. 
However, we emphasise once again that, despite
obvious similarities, the object
$\Delta^{1\,1}_{\mu\nu}(p^2/\Lambda^2)$ is not a
(regularized) Feynman propagator and that at no
stage have we fixed the gauge.
Notice that we can immediately
take $D \rightarrow 4$, since preregularization plays
no role here. Indeed,
this highlights the fact that we only ever
use dimensional regularization as a prescription for
removing finite surface
terms present as a consequence of the 
Pauli-Villars regularization
provided by the $SU(N|N)$ scheme~\cite{SU(N|N)}.
All necessary UV regularization in~\eqs{eq:WL1-nearly}{eq:WL1-nearlyb}
is provided by the cutoff functions buried in the
effective propagator.

Explicitly evaluating the contour integrals 
we find that
\[
	\lim_{T/L \rightarrow \infty} W^{(2)}_R(\bar{\Loop}) = g^2 \frac{C_2(F) T}{4 \pi L},
\]
recovering the standard result.

\section{Conclusion}
\label{sec:Conc}

We have described how to compute
the expectation values of renormalized
gauge invariant operators in a manifestly
gauge invariant way, within the
framework of the exact renormalization
group. The methodology has been illustrated
with a computation of the $\Or (g^2)$
correction to the rectangular Wilson
loop with sides $T \gg L$.

The key elements of the methodology are
as follows. 
Given our regularized $SU(N)$ gauge theory
defined at the bare scale, $\Lambda_0$,
we add a source term $J \GIO_{\Lambda_0}$ for
the gauge invariant operator, $\GIO_{\Lambda_0}$. As
we integrate out modes, so the source
term evolves. Although this generates
a Taylor series in $J$, the only term which
contributes to $\expectation{\GIO}_R$ is the one linear
in $J$ which, after specializing to
the small $J$ limit, we denote by $\GIO_\Lambda$.
\Fig{fig:Oflow} gives the flow of
this component.

We then derived equation~\eq{eq:GIO-zero}, which
states that the expectation value of our gauge
invariant operator is simply given by 
$\GIO_{\Lambda =0}$. Thus, in conjunction,
\figs{fig:Flow}{fig:Oflow} and equation~\eq{eq:GIO-zero}
allow us to compute the expectation value of
an arbitrary gauge invariant operator 
(in some approximation scheme).

The rest of the paper was devoted to 
exploring the formalism in the perturbative
domain. It was here that the considerable
effort invested in~\cite{Thesis,Primer,RG2005,mgiuc}
to understand the structure of perturbative
$\beta$-function coefficients really paid off.
The associated developments
allowed us to directly obtain~\eq{eq:Central},
which gives an extremely compact diagrammatic
expression for the perturbative corrections
to $\expectation{\GIO}_R$. We note that this 
expression makes use of
the diagrammatic function, $\bar{\Q}$,
given by~\eqs{eq:Q}{eq:Qbar}. 
This function
depends only on Wilsonian effective action vertices,
effective propagators and (components of)
gauge remainders. There is no explicit
dependence on either the seed action
or the covariantization of the ERG kernels.

Whilst the perturbative treatment is useful
both to gain experience with the techniques and
also to demonstrate that practical calculations
can be straightforwardly (and correctly) performed,
the real challenge is to apply the formalism
non-perturbatively. Of course, the key results,
shown in \figs{fig:Flow}{fig:Oflow} and equation~\eq{eq:GIO-zero},
are defined non-perturbatively.
The main difficulty is deciding how best
to approximate the flow equation where
there is no obviously small parameter
in which to expand (for speculations on
whether it might be possible to perform
a strong coupling expansion in the inverse
of the \emph{renormalized} coupling see~\cite{conf}).
However, some inspiration for this may be
provided by the perturbative treatment. We know
that
for operators which correspond
to physical observables, the expression
for $\GIO_{\Lambda=0}$ must be universal.
Obviously, such an expression is independent
of the details of the seed action
or the covariantization of the ERG kernels.
Thus, it is natural speculate whether, non-perturbatively,
$\GIO_{\Lambda=0}$ can be written in 
terms of a  generalization
of $\bar{\Q}$; indeed, this generalization has
now been found~\cite{InPrep}. Nevertheless,
this
generalized diagrammatic function possesses
an infinite number of vertices and so much work remains
to be done to extract useful information. However,
this surely represents a desirable, direct starting
point for attacking non-perturbative problems
within the ERG formalism.

\begin{acknowledgments}
I would like to thank Tim Morris for useful
discussions and helpful comments.
I acknowledge financial support from PPARC and IRCSET.
\end{acknowledgments}

\end{document}